\documentclass[11pt,twocolumn,a4paper]{article}

\usepackage{bm,amsmath,amssymb,amsfonts}

\usepackage[dvips]{graphicx}
\usepackage[margin=12pt,font=normalsize,labelfont=bf]{caption}

\usepackage{color}
\definecolor{brown}{rgb}{1,0.5,0}

\usepackage{setspace}
\begin{document}

\onecolumn

\begin{center}
{\bf{\Large\textcolor{brown}{On the extendedness of eigenstates in a 
hierarchical lattice: 
a critical view}}}\\
~\\
\textcolor{blue}{Biplab Pal}\renewcommand{\thefootnote}
{\ensuremath{\fnsymbol{footnote}}}\footnote{{\bf Corresponding Author}: 
Biplab Pal \\
$~$\hspace {0.45cm} Department of Physics,University of Kalyani, \\
$~$\hspace {0.45cm} Kalyani, West Bengal 741 235, India \\
$~$\hspace {0.45cm} Electronic mail: biplabpal2008@gmail.com \\
$~$\hspace {0.45cm} Tel: +91 33 2582 0184\\
$~$\hspace {0.45cm} Fax: +91 33 2582 8282}\textcolor{blue}{, Arunava 
Chakrabarti} \textcolor{blue}{and Nitai Bhattacharya}\\
\it Department of Physics,
University of Kalyani,
Kalyani, West Bengal 741 235, India \\
~\\
{\bf Abstract}
\end{center}

We take a critical view at the basic definition of extended single particle 
states in a non-translationally invariant system. For this, we present the 
case of a hierarchical lattice and incorporate long range interactions that 
are also distributed in a hierarchical fashion. We show that it is possible 
to explicitly construct eigenstates with constant amplitudes (normalized to 
unity) at every lattice point for special values of the electron-energy. 
However, the end-to-end transmission, corresponding to the above energy of 
the electron in such a hierarchical system depends strongly on a special 
correlation between the numerical values of the parameters of the 
Hamiltonian. Keeping the energy and the distribution of the amplitudes 
invariant, one can transform the lattice from conducting to insulating 
simply by tuning the numerical values of the long range interaction. The 
values of these interactions themselves display a fractal character.
\vskip 0.4cm
\begin{flushleft}
\end{flushleft}

\newpage
\section{Introduction}

The problem of localization of single particle states that was initially 
raised and solved by Anderson and others~\cite{anderson} still remains very 
much alive~\cite{fulde,dassarma} and, have given birth to an enormous and 
highly rich literature~\cite{guinea}-\cite{moura2}. The fundamental result 
is that, in one and two dimensions all the single particle eigenstates are 
exponentially localized in presence of uncorrelated 
disorder~\cite{anderson}, while in three dimensions one witnesses the 
possibility of a metal-insulator transition.

Interesting twist in the concept of electron localization in low dimensional 
systems came up with the work of Dunlap et al~\cite{dunlap,kundu} who 
discussed that a short range positional correlation between the constituents 
of a one dimensional disordered chain could lead to resonant (extended) 
eigenstates with high transmittivity. The basic cause was traced back to a 
local resonance in a cluster of impurities, and the case was referred to as 
the {\it random dimer model} (RDM). The idea was tested in low dimensions 
with various form of disorder by many others~\cite{diez}-\cite{hilke}. 
Later, it was argued, based on numerical diagonalization of the Hamiltonian 
that, suitable long range positional correlations between the potentials 
assigned to the atomic sites in a linear chain of atoms one could initiate a 
{\it metal-insulator transition} even in one dimension~\cite{moura1,moura2}.   

The idea of generating {\it extended eigenstates} in low dimensional systems 
without any translational order (disordered systems provide one such class) 
was further extended to the infinite quasi-periodic chains, where one could 
unravel an infinite variety of these unscattered states owing to the self 
similarity of the lattices~\cite{snk,enrique}. Similar studies were carried 
out on fractal networks~\cite{wang1}-\cite{schwalm4} where, even in the 
absence of any local resonating clusters such as in the case of the 
RDM~\cite{dunlap} or in the case of a quasi-periodic chain~\cite{snk} one 
finds an infinite number of extended single particle states. In fact, 
Schwalm and Moritz~\cite{schwalm4} have argued, based on an extensive 
numerical support that, even a continuum of extended eigenstates might exist 
in a class of fractal lattices, a possibility that was also pointed out 
earlier somewhere else~\cite{ac2}.

In all these exciting conceptual developments, a critical issue is 
practically overlooked. This is the question of categorizing an extended 
(resonant) state. More explicitly, let us raise the question, ``when do we 
call an eigenstate {\it extended} ?". An obvious answer is obtained by 
looking at the amplitude of the wave function at a given energy. A non 
trivial distribution of the amplitude throughout an infinite lattice very 
legitimately points to an extendedness of the eigenfunction. A second way to 
characterize an extended wave function is through a calculation of the 
electronic transmission across the lattice at the concerned energy. Usually, 
an extended state should give rise to a ballistic transmission. This second 
criterion is compatible to the first one in all systems where translational 
symmetry prevails. However, its not apparent whether these two demands 
always go hand in hand in systems where extended states exist even in the 
absence of any translational order (such as the fractals for example), and a 
serious introspection of the issue is in order.

In the present communication, we specifically address this problem. 
We provide explicit example of a hierarchical lattice with long range 
interactions where, eigenstates with {\it identical} (and non-zero) 
amplitudes can be constructed at all the lattice points for a special energy 
of the electron. This construction demands a well defined correlation 
between the numerical values of a subset of the Hamiltonian parameters. The 
energy of the electron can be chosen independent of all or, a sub-set of the 
hierarchical long range interactions. It is then shown that the corner-to-
corner propagation of an electron depends crucially on the strengths of the 
long range interactions rendering, in some cases, the lattice completely 
transparent to an incoming electron. In other situations, with a different 
choice of the hierarchical parameter, a topologically identical lattice with 
the same constant distribution of amplitudes of the eigenfunction, becomes 
completely opaque to an electron with the same energy as in the first case. 
This observation, to our mind, introduces in a possible conceptual conflict 
between the extendedness of an eigenstate and a ballistic transmission, 
particularly in such hierarchical systems. 

We discuss two kinds of hierarchical interaction in a fractal network 
designed in the line of the well known Berker lattice~\cite{schwalm3}. 
Hierarchical structures using superconducting wire networks have already 
been fabricated and studied experimentally~\cite{gordon}. With the present 
day nano-technology, tailor-made geometries with quantum dots are also 
possible to fabricate using scanning tunnel microscope(STM) as tweezers. Our 
proposed structures are therefore not far from reality. Also, by controlling 
the proximity of the dots one can induce tunnel hopping almost at will.

We work within a tight binding approach. A real space renormalization group 
(RSRG) decimation scheme~\cite{southern} is employed to calculate the 
corner-to-corner electronic transmission in finite but arbitrarily large 
lattices, and to analyze the character of the electron states. Incidentally, 
a similar analysis has been made recently~\cite{ac3} on a Koch fractal with 
hierarchical interactions, and similar questions have been raised. The 
present lattice is a non-trivial generalization of the Koch 
fractal\cite{maritan}, and exhibits a richer spectrum of results, yet 
maintaining the basic issue, viz., the nature of {\it extended} eigenstates 
in such fractal lattices.

In what follows, we describe the results. In section 2, we describe the 
model and the essential method in resolving the problem. Numerical results 
and related discussion are presented in section 3, and we draw our 
conclusions in section 4.
\section{The model and the method}

\subsection{The Hamiltonian and the decimation scheme}

\begin{figure}[ht]
\centering 
\includegraphics[width=12cm,height=12cm,angle=0]{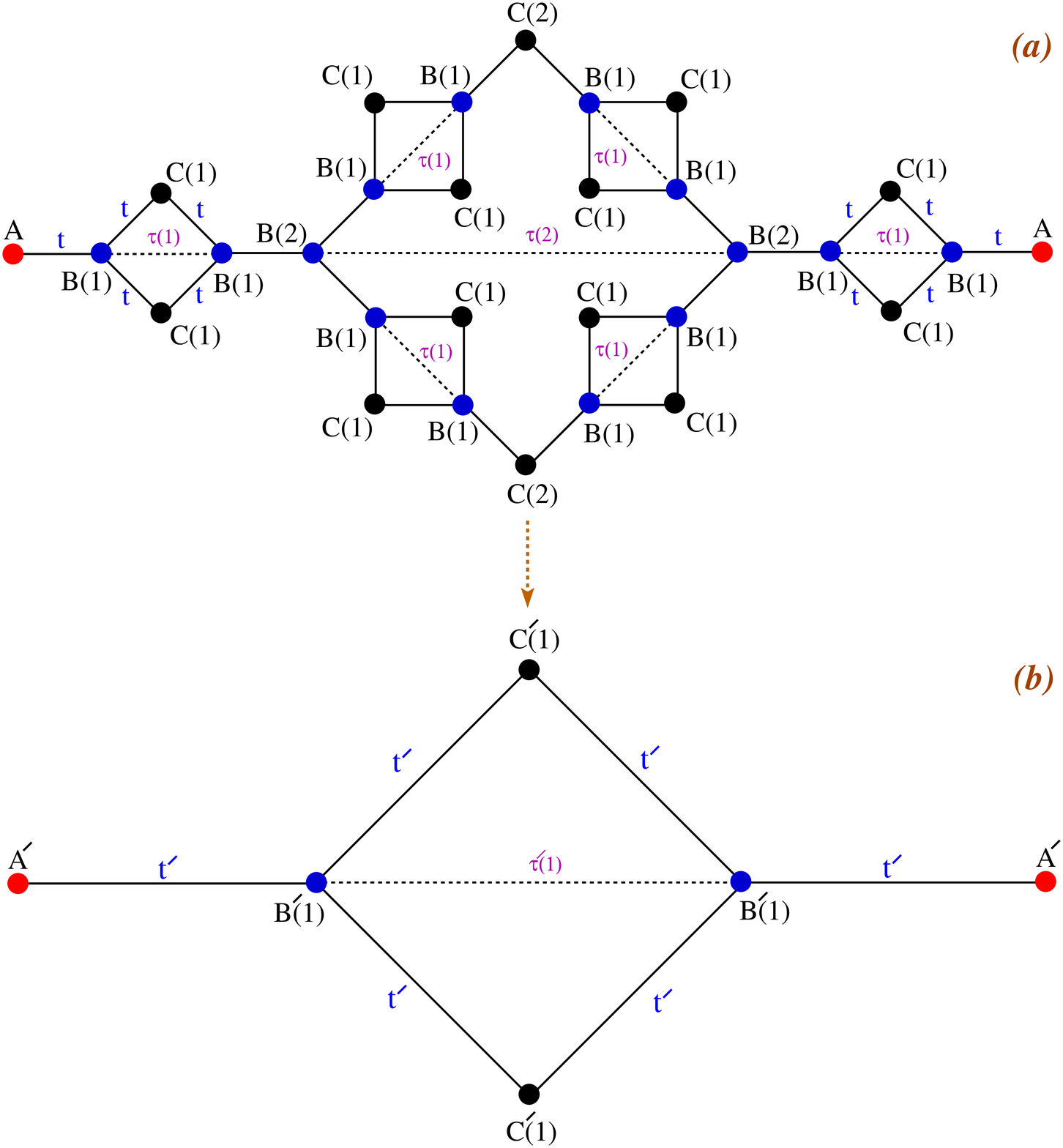}
\caption{(a) Schematic diagram of the 2nd generation of a 
hierarchical lattice with long range interaction along 
the {\it `axial'} direction. {\it `A'} denotes the extreme sites, 
while $B(n)$ and $C(n)$ represent the bulk sites depending on 
their positions in the lattice, $n$ being the hierarchy index. 
{\it `t'}~represents the nearest neighbor hopping integral and 
$\tau(n)$ represents the value of the long range hopping integral 
in the $n$-th level of hierarchy. (b) The renormalized version of (a).}
\label{axial}
\end{figure}

\begin{figure}[ht]
\centering 
\includegraphics[width=12cm,height=12cm,angle=0]{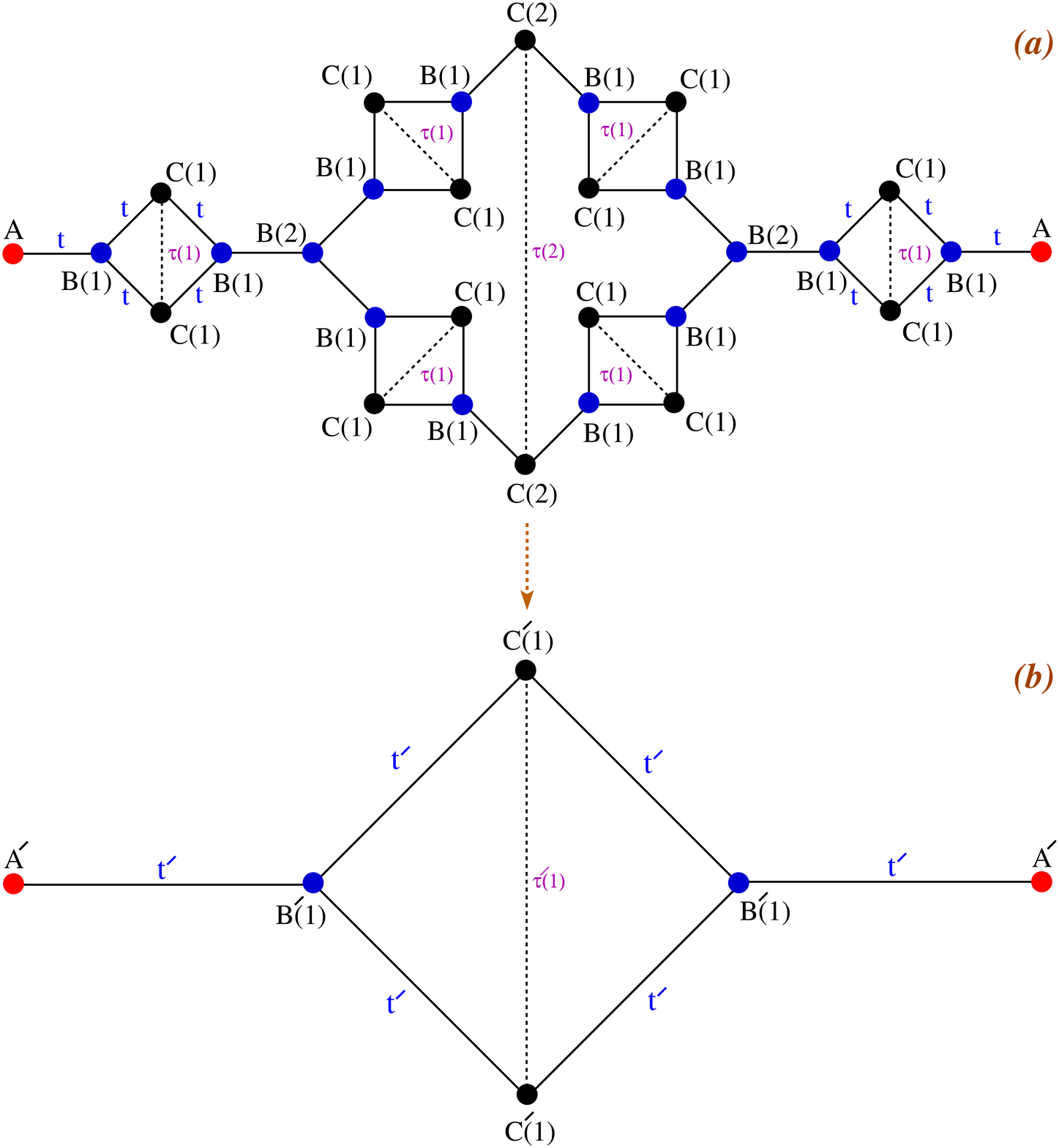}
\caption{(a) Schematic diagram of the 2nd generation of a 
hierarchical lattice with long range interaction along 
the {\it `transverse'} direction. {\it `A'} denotes the extreme sites, 
while $B(n)$ and $C(n)$ represent the bulk sites depending on 
their positions in the lattice, $n$ being the hierarchy index. 
{\it `t'}~represents the nearest neighbor hopping integral and 
$\tau(n)$ represents the value of the long range hopping integral 
in the $n$-th level of hierarchy. (b) The renormalized version of (a).}
\label{transverse}
\end{figure}

We begin by referring to Fig.~\ref{axial} and Fig.~\ref{transverse}. We have 
designed a hierarchical lattice with two different configurations, in one of 
which the long range interaction is along the axial direction 
(Fig.~\ref{axial}) and in the other one, it is along the transverse direction 
(Fig.~\ref{transverse}). We adopt a tight binding formalism, and incorporate 
only the nearest neighbor hopping. The tight binding Hamiltonian for non-
interacting electrons in the Wannier basis is given by, 
\begin{equation}
H=\sum_{i}\epsilon_{i}|i\rangle\langle{i}|
+\sum_{\langle ij \rangle} \:[\;t_{ij}|i \rangle \langle{j}| + 
t_{ji}|j \rangle \langle{i}|\;]
\label{Hamiltioan}
\end{equation}
where, $\epsilon_{i}$ is the on-site potential of an electron on the $i$-th 
atomic site and $t_{ij}=t_{ji}$ is the nearest neighbor hopping integral 
between the $i$-th and $j$-th sites. For the nearest neighboring sites of 
the lattice we assume $t_{ij}=t$, while the long range hopping integrals are 
$t_{ij}=\tau$, and are assumed to follow a particular rule: 
$\tau(n)=\lambda^{n}t$ where, $n$ represents the level of hierarchy, and 
$\lambda$ is the hierarchy parameter. For such a hierarchical lattice of 
finite but arbitrarily large size, the on-site potential $\epsilon_{i}$ will 
be assigned values $\epsilon_{A}$ for the two extreme sites and 
$\epsilon_{B(n)}$, $\epsilon_{C(n)}$ etc. for the bulk sites depending on 
their positions on the lattice as explained in Fig.~\ref{axial} and 
Fig.~\ref{transverse}.

Exploiting to the self similarity of the lattice, we can apply the real 
space renormalization group (RSRG) decimation scheme~\cite{southern} to 
decimate an appropriate subset of sites. In decimating those subset of sites, 
we have used the standard difference equation, which is an equivalent 
description of the Schr\"{o}dinger equation, viz.,
\begin{equation}
(E-\epsilon_{i})\:\psi_{i}=\sum_{j}\;t_{ij}\:\psi_{j}
\label{diffeqn}
\end{equation}
Here, $\psi_{i}$ is the amplitude of the wave function at the $i$-th atomic 
site and the index $j$ represents the nearest neighbors of $i$. The 
recursion relations for the site energies and hopping integrals for the two 
cases are given by: 

\begin{center}
{\bf I. The axial case} 
\end{center}

\begin{eqnarray}
\epsilon_{A}' &=& \epsilon_{A} + \dfrac{\alpha t^{2}}
{\beta(1-\gamma^{2})} \nonumber \\
\epsilon_{B(n)}' &=& \epsilon_{B(n+1)} + \dfrac{3\alpha t^{2}}
{\beta(1-\gamma^{2})} \nonumber \\
\epsilon_{C(n)}' &=& \epsilon_{C(n+1)} + \dfrac{2\alpha t^{2}}
{\beta(1-\gamma^{2})} \nonumber \\
t' &=& \dfrac{\alpha \gamma t^{2}}
{\beta(1-\gamma^{2})} \nonumber \\
\tau'(n) &=& \tau(n+1)\qquad \text{for all}\ n~\geq 1
\label{rgrelations1}
\end{eqnarray}

\begin{equation}
\text{where},\quad \alpha = E-\epsilon_{C(1)},\ 
\beta = \alpha (E-\epsilon_{B(1)})-2t^{2}\ 
\text{and}\ \gamma = [\,2t^{2}+\alpha \tau(1)\,]\,/\beta 
\end{equation}

\begin{center}
{\bf II. The transverse case} 
\end{center}

\begin{eqnarray}
\epsilon_{A}' &=& \epsilon_{A} + \dfrac{\delta \mu t^{2}}
{(\mu^{2}-4t^{4})} \nonumber \\
\epsilon_{B(n)}' &=& \epsilon_{B(n+1)} + \dfrac{3\delta \mu t^{2}}
{(\mu^{2}-4t^{4})} \nonumber \\
\epsilon_{C(n)}' &=& \epsilon_{C(n+1)} + \dfrac{2\delta \mu t^{2}}
{(\mu^{2}-4t^{4})} \nonumber \\
t' &=& \dfrac{2\delta t^{4}}
{(\mu^{2}-4t^{4})} \nonumber \\
\tau'(n) &=& \tau(n+1)\qquad \text{for all}\ n~\geq 1
\label{rgrelations2}
\end{eqnarray}

\begin{equation}
\text{where},\quad \delta = E-\epsilon_{C(1)}-\tau(1)\ 
\text{and}\ \mu = \delta [E-\epsilon_{B(1)}]-2t^{2}
\end{equation}
We present our numerical results in the next section using the above set of recursion relations. 

\subsection{The transmission coefficient}

To get the end-to-end transmission coefficient of an $\ell$-th generation 
fractal, we clamp the system between two semi-infinite ordered leads. The 
leads, in the tight binding model, are described by a constant on-site 
potential $\epsilon_{0}$ and a nearest neighbor hopping integral $t_{0}$. We 
then renormalize the lattice $\ell$ times to reduce it into an effective 
dimer consisting of the `renormalized' $A$-atoms, each having an on-site 
potential equal to $\epsilon_A^{(\ell)}$ and the effective $A\text{--}A$ 
hopping integral $t^{(\ell)}$. The transmission coefficient is then obtained 
by the well known formula~\cite{stone}, 
\begin{equation}
T=\dfrac{4\sin^{2}ka}{\left[(P_{12}-P_{21})+(P_{11}-P_{22})\cos ka 
\right]^{2}+\left[(P_{11}+P_{22})\sin ka \right]^{2}}
\label{trcoeff}
\end{equation}
where, 
\begin{eqnarray}
P_{11} &=& \dfrac{[E-\epsilon_{A}^{(\ell)}]^{2}}
{t_{0}t^{(\ell)}}-\dfrac{t^{(\ell)}}{t_{0}} \nonumber \\
P_{12} &=& -\dfrac{[E-\epsilon_{A}^{(\ell)}]}{t^{(\ell)}} \nonumber \\
P_{21} &=& -P_{12} \nonumber \\
P_{22} &=& -\dfrac{t_{0}}{t^{(\ell)}}
\label{matrix}
\end{eqnarray}
\begin{equation*}
\text{Here},\ \cos ka = (E-\epsilon_{0})/2t_{0}
\end{equation*}
$a$ is the lattice spacing in the leads, and throughout the calculation we shall set $\epsilon_{0}=0$, and $t_{0}=1$.
\section{Numerical results and discussion}

\subsection{The eigenvalue spectrum}

\begin{figure}[ht]
\centering 
\includegraphics[width=10cm,height=15cm,angle=-90]{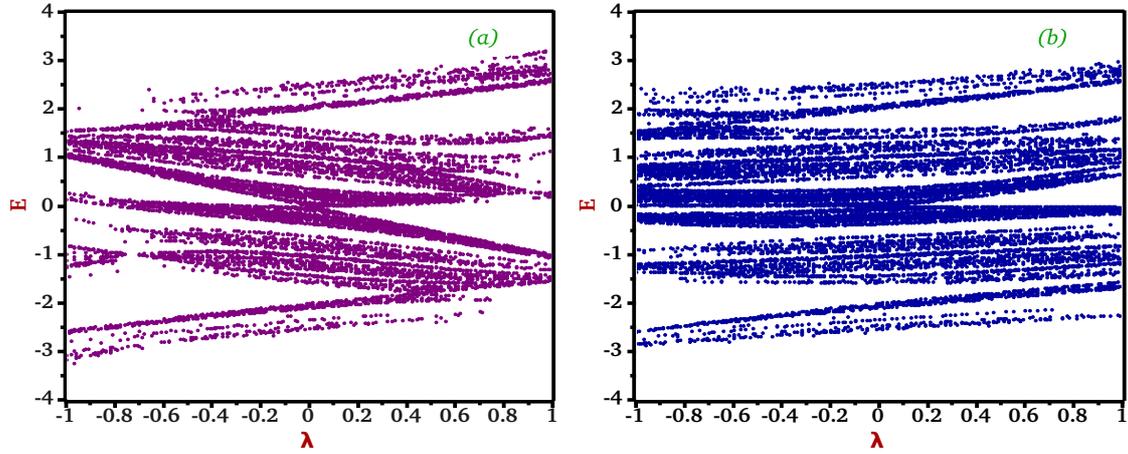}
\caption{Energy eigenvalue spectrum of a hierarchical fractal lattice as a 
function of the hierarchy parameter $\lambda$, obtained from the trace of 
the transfer matrix for the $7$th generation fractal, taken as the `unit 
cell'. We have set $\epsilon_{A} = \epsilon_{B(n)} = \epsilon_{C(n)} = 0$ 
with $n=7$ and $t=1$. (a) The axial case and (b) The transverse case.}
\label{spectrum}
\end{figure}

Before discussing the precise point of interest, we prefer to have a look at 
the eigenvalue spectra of the proposed hierarchical structures. In 
particular, we examine the formation of the bands and the gaps as a function 
of the hierarchy parameter $\lambda$ both for the axial and the transverse 
cases. To this end, we use a well known trick used frequently in the case of 
a quasi-periodic lattice~\cite{kkt}, and later used for hierarchical 
structures as well~\cite{ac3,schneider}. We sequentially generate periodic 
approximants of the original hierarchical structure, calculate the trace of 
the transfer matrix corresponding to a `unit cell' and extract those values 
of energy for which the magnitude of the trace of the transfer matrix 
remains bounded by $2$~\cite{kkt}. The results are shown for the axial and 
the transverse cases in Figs.~\ref{spectrum}(a) and (b) respectively. The 
common feature in both the figures is the presence of multiple bands and 
gaps. A variation of the hierarchy parameter $\lambda$ leads to band 
overlapping. Specifically speaking, in the axial case, the density of 
allowed energy values is larger in the range $-0.2t \leq ~\lambda ~\leq 
0.2t$. Band crossings maximize in this area. An increase in the numerical 
value of $\lambda$ leads to a thinning of the spectrum. Influence of 
$\lambda$ on the spectrum of the transverse model is also similar.   

\subsection{The unusual states}

We now examine some special situation which is the focus of this article.

\vskip .25in
\noindent
{\bf \emph {(a) The axial model}}
\vskip .25in
Let us begin with the axial model. It is easy to verify that, if we 
choose the energy of the electron $E = \epsilon_A + t$, then one can 
construct, by hand, a wave 
function with amplitude equal to {\it unity} at every lattice point, 
provided one also fixes 
\begin{eqnarray}
\epsilon_{B(n)} + \tau(n) & = & \epsilon_{A} - 2 t \nonumber \\
\epsilon_{C(n)} & = & \epsilon_{A} - t \qquad \text{for all}\ n~\geq 1
\label{correlation1}
\end{eqnarray}
Two points are worth noting. First, the distribution of  amplitude thus 
constructed is independent of the individual values of $\epsilon_{B(n)}$ and 
$\tau(n)$, and only requires the special correlation in their numerical 
values as suggested in Eq.~\eqref{correlation1}. Thus, in principle, 
$\epsilon_{B(n)}$ can be chosen out of any sequence of uncorrelated random 
numbers. The values of the hopping integrals $\tau(n)$ only then need to be 
selected accordingly. Thus the constructed amplitude distribution remains 
valid even for a random choice of a subset of the on-site potentials 
$\epsilon_{B(n)}$, and presents a new kind of {\it extended} wave function. 
However, this construction can not automatically be related to the  good 
transmission property of any large but finite lattice. This is easily 
understood when one appreciates that the choice of the energy didn't depend 
on the hierarchically distributed hoppings viz., $\tau(n)$, but the end-to-
end transmission is bound to be sensitive to the individual values of 
$\tau(n)$. 

\begin{figure}[ht]
\centering 
\includegraphics[width=10cm,height=15cm,angle=-90]{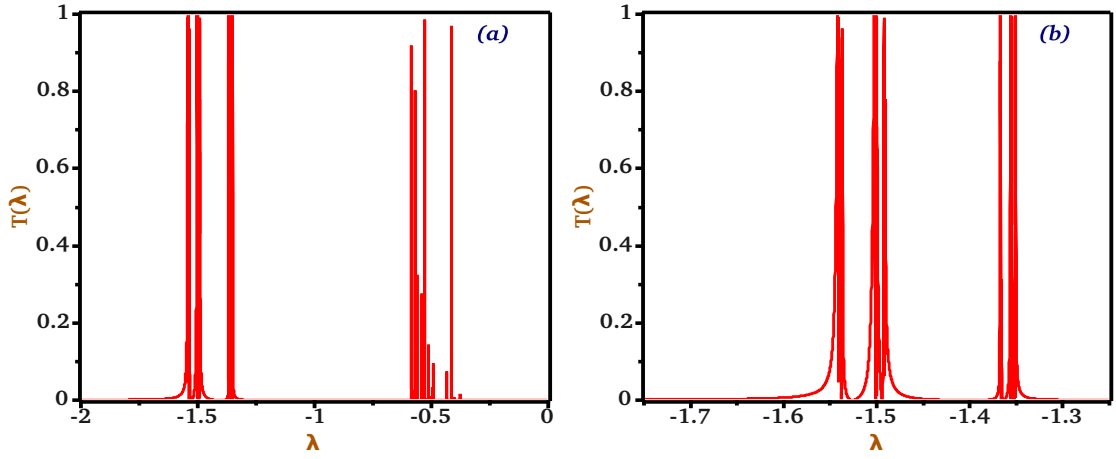}
\caption{(a) Transmission coefficient across a $7$th generation fractal 
network (the axial case)  
for $E = \epsilon_A + t$ with $\epsilon_{B(n)} + \tau(n) = 
\epsilon_A - 2 t$, and $\epsilon_{C(n)} =\epsilon_A - t$. (b) Fine scan of a 
selected part of (a) to reveal the self-similar distribution of $\lambda$. 
We have set $\epsilon_A = 0$, and $t = 1$.}
\label{transm1}
\end{figure}

This is illustrated in Fig.~\ref{transm1}(a), where we have worked out the 
transmission coefficient across a $7th$ generation hierarchical network and 
for $\tau(n) = \lambda \tau(n-1)$, with $\tau(1) = \lambda t$. Energy is set 
at $E = \epsilon_A + t$. The corresponding eigenstate has equal amplitude 
(normalized to unity) at all lattice points. The transmission spectrum shows 
that the $7th$ generation network is transparent to an incoming electron 
with the above energy only when $\lambda$ assumes a specific set of values. 
The transmission in this range of $\lambda$ is completely ballistic for 
certain values. We may assign the wave function in this case  the status of 
an {\it extended state}. Interestingly, the same distribution of the 
amplitudes of the wave function, at the same energy $E = \epsilon_A + t$ 
makes the lattice completely opaque to the incoming electron for other 
ranges of $\lambda$. The energy definitely corresponds to an eigenstate of 
the system, as has been verified by calculating the local density of states 
at the sites of an infinite hierarchical lattice for the axial case. Also, 
the very fact that, one is able to construct such a state on a lattice, no 
matter how large it is, automatically confirms that it is an eigenstate. 
This observation leads to the question of a proper identification of an {\it 
extended state} in a non-translationally invariant system.

\begin{figure}[ht]
\centering 
\includegraphics[width=10cm,height=15cm,angle=-90]{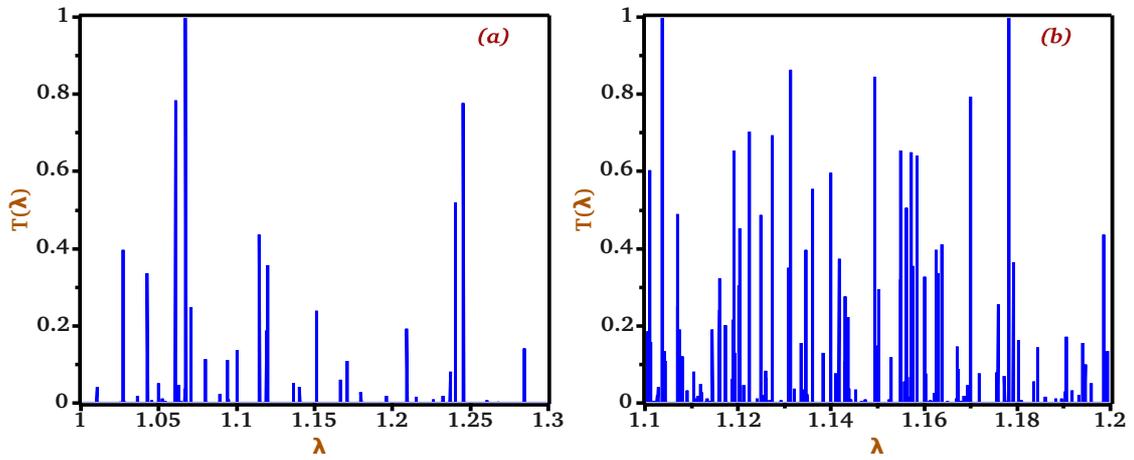}
\caption{(a) Transmission coefficient across a $7$th generation fractal 
network (the axial case) for $E = \epsilon_A' + t'$ and (b) a fine scan of a selected part of 
(a) to reveal the self-similar distribution of $\lambda$. We have set 
$\epsilon_{A}=\epsilon_{B(1)}=\epsilon_{C(1)}=0$, and $t=\tau(1)=1$.}
\label{transm2}
\end{figure}

The procedure can be implemented on say, a one step renormalized lattice.
That is, we can extract an energy eigenvalue by solving the equation 
$E = \epsilon_A' + t'$. The energy turns out to be a function of 
$\epsilon_A$, $t$, $\epsilon_{C(1)}$, $\epsilon_{B(1)}$ and $\tau(1)$, but 
remains independent of $\epsilon_{B(n)}$, $\epsilon_{C(n)}$ and $\tau(n)$ 
for $n \geq 2$. Once again we are to choose,
\begin{eqnarray}
\epsilon_{B(n)} + \tau(n) & = & \epsilon_{A}' - 2 t' \nonumber \\
\epsilon_{C(n)} & = & \epsilon_{A}' - t' \qquad \text{for all}\ n~\geq 2
\label{correlation2}
\end{eqnarray}
The entire scheme works as before. In Fig.~\ref{transm2}(a) we illustrate 
the transmission coefficient for $E = 0$. This energy is extracted by 
solving the equation $E = \epsilon_A' + t'$, where we have chosen 
$\epsilon_{A}=\epsilon_{B(1)}=\epsilon_{C(1)}=0$, and $t=\tau(1)=1$. A fine 
scan of a selected portion of Fig.~\ref{transm2}(a) is presented in 
Fig.~\ref{transm2}(b) to show the self similar distribution of the values of 
$\lambda$. The process can, in principle, be continued and one can extract 
energy eigenvalues for such {\it unusual} eigenstates by solving the 
equation $E = \epsilon_{A}^{(\ell)} + t^{(\ell)}$ at any $\ell$-th stage of 
renormalization. Ideally, we thus have an infinite number of such 
eigenstates.

\vskip .25in
\noindent
{\bf \emph{(b) The transverse model}}
\vskip .25in
For the transverse model, we can use an identical string of arguments to 
construct an eigenstate with an amplitude equal to {\it unity} at every 
lattice point. For this construction we fix the energy $E = \epsilon_A + t$, 
and demand,
\begin{eqnarray}
\epsilon_{B(n)} & = & \epsilon_{A} - 2 t \nonumber \\
\epsilon_{C(n)} + \tau(n) & = & \epsilon_{A} - t \qquad \text{for all}\ 
n~\geq 1
\end{eqnarray}
This choice of parameters leads to a completely new scenario in comparison 
to the axial case. Here, for values of the hierarchy 
parameter $\lambda>1$, it is found that at any $\ell$-th stage of 
renormalization the on-site potential at the $C$-sites grow following the 
rule: 
\begin{equation} 
\epsilon_{C(n)}^{(\ell)} = 
\lambda \epsilon_{C(n-1)}^{(\ell)} - \xi^{(\ell)}(\lambda)
\end{equation}
where, $\xi^{(\ell)}(\lambda)$ is a constant, function of $\lambda$, 
and the on-site potential at the $B$-sites at any $\ell$-th stage of 
renormalization $\epsilon_{B(n)}^{(\ell)}$ remains constant. Thus an 
incoming electron with energy $E = \epsilon_A + t$ will face effectively 
higher and higher potential barriers, offered by the `$C$' sites while 
travelling through the lattice. This will lead to a gradual decay 
of the end-to-end transmission as the system grows in size. In the 
thermodynamic limit the hierarchical lattice will remain non-conducting. 
For $\lambda \leq1$, no regular pattern in the on-site potentials is 
observed. However, the hopping integral $t$ always decays to zero for
$E=\epsilon_A+t$.

We have also examined the case, 
where the energy is extracted from a one step renormalized lattice by 
solving the equation $E = \epsilon_A' + t'$. For this we are free to choose 
$\epsilon_A$, $\epsilon_{B(1)}$, $\epsilon_{C(1)}$, $t$ and $\tau(1)$ 
freely, that is, in an uncorrelated fashion. The correlation now sets in 
from the hierarchy level $n = 2$ onwards. For example, we have examined the 
special situation when $\epsilon_A = \epsilon_{B(1)} = \epsilon_{C(1)} = 0$, 
and $t = \tau(1) = 1$. The roots are, $E = -1.90321$, $0.193937$, and 
$2.70928$. The transmission in each case drops fast as the lattice grows in 
size. The amplitude-distribution on a one step renormalized lattice is such 
that $\psi_i = 1$ on every vertex of the {\it renormalized lattice}. 
Therefore, the states are still strictly localized. Thus we can say that, a 
hierarchically distributed long range hopping in the transverse direction does not allow the lattice to 
be conducting. 

Thus, once we appreciate that a long range tunnel hopping can be associated 
with the proximity of the atoms in the structure, we see that a proximity 
along the principal axis of the fractal allows both for conduction and 
localization, whereas, a proximity in the transverse direction makes the 
lattice non-conducting in general.

Before we end, it should be emphasized that, all these discussions are made 
with reference to special values of the energy $E$ for which one can 
construct a unique spatially extended distribution of the eigenfunctions. 
However, hierarchical lattices, as already discussed in the introduction, 
may possess both localized and extended eigenstates coexisting in the 
spectrum. For example, we have worked out a special situation in the 
transverse case, where a different set of values for the Hamiltonian 
parameters and the electron-energy may lead to a one cycle fixed point of 
the parameter space. The eigenstate in this particular case is definitely 
extended and the transport is high. However, we refrain from entering into 
this aspect to save space.
\section{Conclusions}

In conclusion, we have undertaken a detailed study of the electronic states 
and transport across a hierarchical lattice corresponding to a special set 
of energy eigenvalues. The central issue has been to examine if 
an unambiguous definition of the 
extendedness of an eigenstate in a lattice that lacks translational 
symmetry is obtainable. We come to the conclusion that, a state that looks 
{\it extended} by construction is not necessarily conducting, and the 
mobility of the state is strongly sensitive to a correlated choice of a 
subset of the system parameters. Even within the same basic lattice 
topology, a long range correlated tunnel hopping along the principal axis is 
found to lead to both extended and localized states depending on the value 
of the hierarchy parameter. The hierarchically long range hopping in the 
transverse direction, on the other hand, makes the eigenfunction localized 
in the lattice.
\vskip 0.3in
\noindent
{\bf\small ACKNOWLEDGMENT}
\vskip 0.2in
\noindent
One of the authors (B. Pal) acknowledges financial assistance 
from DST, India through the INSPIRE fellowship.
\newpage

\newpage
\noindent
{\bf\large Figure Captions:}
\vskip 0.3in

\noindent
{\bf Figure 1:} (Color online). (a) Schematic diagram of the 2nd generation of a 
hierarchical lattice with long range interaction along 
the {\it `axial'} direction. {\it `A'} denotes the extreme sites, 
while $B(n)$ and $C(n)$ represent the bulk sites depending on 
their positions in the lattice, $n$ being the hierarchy index. 
{\it `t'}~represents the nearest neighbor hopping integral and 
$\tau(n)$ represents the value of the long range hopping integral 
in the $n$-th level of hierarchy. (b) The renormalized version of (a).
\vskip 0.3in

\noindent
{\bf Figure 2:} (Color online). (a) Schematic diagram of the 2nd generation of a 
hierarchical lattice with long range interaction along 
the {\it `transverse'} direction. {\it `A'} denotes the extreme sites, 
while $B(n)$ and $C(n)$ represent the bulk sites depending on 
their positions in the lattice, $n$ being the hierarchy index. 
{\it `t'}~represents the nearest neighbor hopping integral and 
$\tau(n)$ represents the value of the long range hopping integral 
in the $n$-th level of hierarchy. (b) The renormalized version of (a).
\vskip 0.3in

\noindent
{\bf Figure 3:} (Color online). Energy eigenvalue spectrum of a hierarchical 
fractal lattice as a function of the hierarchy parameter $\lambda$, obtained 
from the trace of the transfer matrix for the $7$th generation fractal, 
taken as the `unit cell'. We have set $\epsilon_{A} = \epsilon_{B(n)} = 
\epsilon_{C(n)} = 0$ with $n=7$ and $t=1$. (a) The axial case and (b) The 
transverse case.
\vskip 0.3in

\noindent
{\bf Figure 4:} (Color online). 
(a) Transmission coefficient across a $7$th generation fractal 
network (the axial case)  
for $E = \epsilon_A + t$ with $\epsilon_{B(n)} + \tau(n) = 
\epsilon_A - 2 t$, and $\epsilon_{C(n)} =\epsilon_A - t$. (b) Fine scan of a 
selected part of (a) to reveal the self-similar distribution of $\lambda$. 
We have set $\epsilon_A = 0$, and $t = 1$.
\vskip 0.3in

\noindent
{\bf Figure 5:} (Color online). (a) Transmission coefficient across a $7$th generation fractal 
network (the axial case) for $E = \epsilon_A' + t'$ and (b) a fine scan of a selected part of 
(a) to reveal the self-similar distribution of $\lambda$. We have set 
$\epsilon_{A}=\epsilon_{B(1)}=\epsilon_{C(1)}=0$, and $t=\tau(1)=1$.
\newpage
\vskip 1in
\noindent
{\bf\large Figures:}

\newpage
\vskip 1in

\newpage
\vskip 1in

\end{document}